\documentclass[12pt]{iopart}
\usepackage{graphicx}
\usepackage{latexsym}
\usepackage{amsthm}

\usepackage{iopams}
\usepackage{url}
\def\aj{AJ  }
\def\apj{ApJ\,  }
\def\apjs{ApJS  }

\def\mnras{MNRAS\,  }

\def\h0units{\mathrm{km\,s^{-1}\,Mpc^{-1}}}
\def\sun{\hbox{$\odot$}}
\begin{document}
\title
{
The Photometric Maximum in  Cosmicflows-2 }
\vskip  1cm
\author     {Lorenzo Zaninetti}
\address    {
Physics Department,
 via P.Giuria 1,\\ I-10125 Turin,Italy 
}
\ead {zaninetti@ph.unito.it}

\begin {abstract}
Based on well known photometric rules for the number of galaxies 
as function  of the distance in Mpc we model
the so called "Great Wall" which is visible 
on the  Cosmicflows-2 catalog.
The gravitational field is evaluated  
at the light of the shell theorem 
and a finite value for the gravitational field 
is numerically  derived.
\end{abstract}
\vspace{2pc}
\noindent{\it Keywords}:
{
Cosmology;
Observational cosmology;
Distances, redshifts, radial velocities, spatial distribution of
galaxies;
Magnitudes and colors, luminosities
}

\section{Introduction}

Before to start 
we briefly review 
the Hubble law, after 
\cite{Hubble1929},
which is a linear relationship 
between expansion velocity  
and distance in 
\begin{equation}
V= H_0 D  = c \, z  
\quad  ,
\label{clz}
\end{equation}
where $H_0$ is the Hubble constant 
 $H_0 = 100 h \mathrm{\ km\ s}^{-1}\mathrm{\ Mpc}^{-1}$, with $h=1$
when  $h$ is not specified,
$D$ is the distance in $Mpc$,
$c$ is  the  light velocity  and
$z$   
is the redshift.
As an example  a  recent  evaluation,
see \cite{Tully2013},
quotes
\begin{equation}
H_0 =(74.4 \pm 3 ) \h0units
\quad.
\end{equation}
The original Hubble's law was based 
on $\approx$ 20 galaxies  of which was known the distance and
the velocity; 
conversely
in the last years  the number of catalogs 
for galaxies with redshift available for public downloading
has progressively grown.
This huge  amount of data allows setting  up  critical tests  
between different theoretical models
over  the various aspects  of the Large Scale Structures.
The layout of the paper is as follows.
In Section \ref{sec_catalogs}, 
we describe some representative  catalogs of galaxies.
In Section \ref{sec_lf}, 
we introduce the adopted luminosity function for galaxies.
In Section \ref{sec_maximum},  we model the 
maximum in the number of galaxies as function of the 
distance.
In Section \ref{sec_gravitational},  we evaluate the gravitational
field as produced by the visible galaxies 
of 
Cosmicflows-2 catalog.
We conclude in Section \ref{sec_conclusions}.

\section{The catalogs of galaxies}
\label{sec_catalogs}

\begin{figure*}
\begin{center}
\includegraphics[width=6cm]{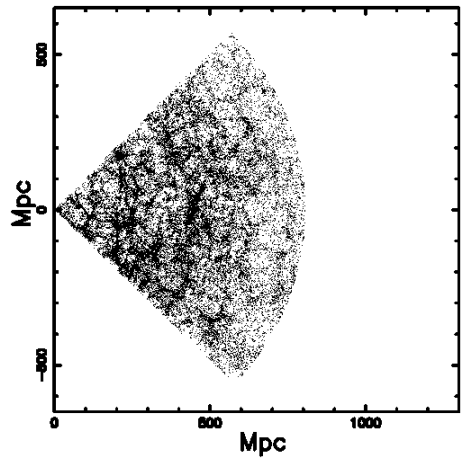}
\end{center}
\caption{Cone-diagram  of the  galaxies  
in the 2dFGRS  with distance  $\in$  $\bigl [4Mpc,470Mpc\bigr ]$.  
This plot contains  117293  galaxies.
}
          \label{slice2dfone}%
    \end{figure*}

\begin{figure}
\begin{center}
\includegraphics[width=6cm]{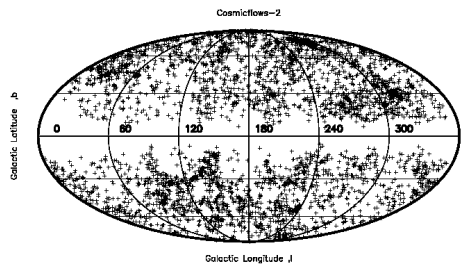}
\end{center}
\caption
{Sky distribution of 4970  Cosmicflows-2's galaxies   
in Galactic coordinates projected using the Mollweide projection.
}
\label{cosmicflows2_mollweide}
\end{figure}

A first kind of  catalog for  galaxies 
is represented by those focused  as  slice   
as  the two-degree Field Galaxy Redshift Survey,
in the following 2dFGRS, see \cite{Colless2001},
or the Sloan Digital Sky Survey (SDSS), see \cite{Berlind2006}.
A second  classification is about 
the all-sky catalogs  such as  
the  2MASS Redshift Survey  (2MRS), see \cite{Huchra2012},
or
the Cosmicflows-2, see \cite{Tully2013}.
As a first  example  
a  strip of  the 2dFGRS is   
shown in Figure~\ref{slice2dfone}.
The above Figure, which is a 3~$^{\circ}$ slice,
shows that the galaxies resides on filaments 
rather to be distributed in a uniform 2D  way.
A first extrapolation allows to state
that the galaxies are disposed 
on the surface of bubbles rather than to fill
uniformly the 3D space.
The statistics of the voids is a topic 
of research and an important 
parameter is the average radius of the 
voids, $<R>$, :
\cite{Vogeley2012} quotes 
$<R> = 18.23 h^{-1}\, Mpc$  
and  Table 1 in \cite{Mao2017} quotes
a variable average radius 
ranging from $<R> = 15.964 h^{-1}\, Mpc$  
to $<R> = 63.347 h^{-1}\, Mpc$ 
according to the selected sample.  
As  second  example  is represented by 
the  overall  
Cosmicflows-2 catalog, data available at  
\url{http://vizier.u-strasbg.fr/viz-bin/VizieR},
and Figure~\ref{cosmicflows2_mollweide} 
reports  the sky distribution of  such a catalog
in Galactic coordinates.

\begin{figure}
\begin{center}
\includegraphics[width=6cm]{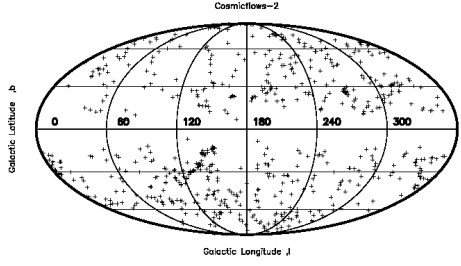}
\end{center}
\caption
{Sky distribution of 713  Cosmicflows-2's galaxies
with distance $\in \bigl[ 63.8 Mpc,83.8 \bigr] Mpc$.
}
\label{cosmicflows2_sele}
\end{figure}

Also in this case  the voids are can be visualized selecting a 
thin (20 Mpc) spherical  shell , see Figure~\ref{cosmicflows2_sele}.

\section{Luminosity Function for Galaxies}
\label{sec_lf}
The distance  modulus is
\begin{equation}
m - M =  5\,\log (d) -5
\quad  ,
\label{distancemodulusgalaxies}
\end{equation}
where
$m$ is the  apparent magnitude,
$M$ is the  absolute magnitude 
and  $d$ is the distance  in pc.

Let $L$, the  luminosity of a galaxy,
be defined  in  $\bigl [0, \infty\bigr ]$.
The Schechter LF of galaxies ,$\Phi$,
see \cite{schechter}, is  

\begin{equation}
\Phi (L;\Phi^*,\alpha,L^*) dL =
(\frac{\Phi^*}{L^*}) (\frac{L}{L^*})^{\alpha}
\exp \bigl ({- \frac{L}{L^*}} \bigr ) dL \quad,
\label{lf_schechter}
\end{equation}
where $\alpha$ sets the slope for low values
of $L$,
$L^*$ is the
characteristic luminosity, and $\Phi^*$ represents
the number of galaxies  per $Mpc^3$.
The  normalization is
\begin{equation}
\int_0^{\infty} \Phi (L;\Phi^*,\alpha,L^*) dL  =
\rm \Phi^*\, \Gamma \left( \alpha+1 \right)
\quad  ,
\label{norma_schechter}
\end{equation}
where
\begin{equation}
\rm \Gamma \, (z )
=\int_{0}^{\infty}e^{{-t}}t^{{z-1}}dt
\quad ,
\end{equation}
is the Gamma function.
The average luminosity,
${\langle L \rangle } $, is
\begin{equation}
{\langle (\Phi (L;\Phi^*,\alpha,L^*) \rangle}
=
\rm L^* \,{\rm \Phi^*  }\,\Gamma \left( \alpha+2 \right)
\quad.
\label{ave_schechter}
\end{equation}
An equivalent form  in absolute magnitude
of the Schechter LF
is
\begin{eqnarray}
\Phi (M;\Phi^*,\alpha,M^*)dM=
\nonumber\\
0.921 \Phi^* 10^{0.4(\alpha +1 ) (M^*-M)}
\exp \bigl ({- 10^{0.4(M^*-M)}} \bigr)  dM \, ,
\label{lfstandard}
\end{eqnarray}
where $M^*$ is the characteristic magnitude.

\begin{figure}
\begin{center}
\includegraphics[width=7cm]{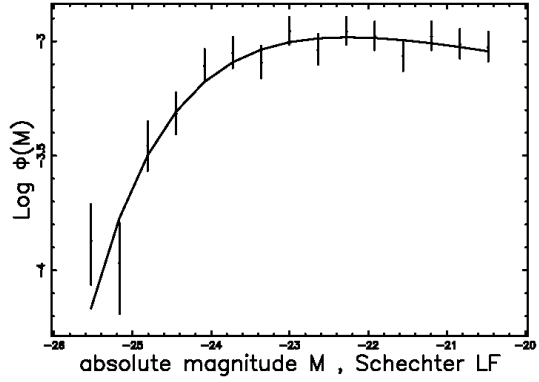}
\end{center}
\caption{The  observed  Schechter  LF for galaxies of
Cosmicflows-2, empty stars with error bar,
and the fit  by  the Schechter LF
when  $Phi^*= 0.0018\,h^3 \,Mpc^{-3}$,
$M^*= -24.2 -5 Log_{10} (h)$
and  $\alpha=-0.83$.
}
 \label{cosmicflows2_lf_fit}%
\end{figure}

A typical result of  the Schechter LF in the case 
of  Cosmicflows-2
is  reported  in Figure~\ref{cosmicflows2_lf_fit}.

\section{The photometric maximum}
\label{sec_maximum}
The flux, $f$,
is
\begin{equation}
f  = \frac{L}{4 \pi r^2}
\quad ,
\label{flux}
\end{equation}
where $r$ is the distance and $L$ the luminosity of the galaxy.
The joint distribution in distance, {\it r},
and  flux,{\it f},  for the number of galaxies is
\begin{equation}
\frac{dN}{d\Omega dr df} = \frac{1}{4 \,\pi}
\int_0^{\infty } 4 \pi r^2  dr
\Phi(\frac{L}{L^*}) \delta (f-\frac{L}{4\,\pi\,r^2} )
\quad ,
\label{nldef}
\end{equation}
were the factor  ($\frac{1}{4 \pi}$)
converts the number density
into density for solid angle
and the Dirac delta function  selects  the required flux.
In the case of 
Schechter LF of galaxies
the number of galaxies as function of the distance  is
\begin{equation}
\frac{dN}{d\Omega dr df} =
\frac{1}{L^*}\,
4\,\pi\,{r}^{4}{\it \Phi^*}\, \left( 4\,{\frac{\pi\,f{r}^{2}}{{\it L^*}}
} \right) ^{\alpha}{{\rm e}^{-4\,{\frac{\pi\,f{r}^{2}}{{\it L^*}}}}
}
\quad.
\label{nfunctionrschechter}
\end{equation}

\begin{figure}
\begin{center}
\includegraphics[width=6cm]{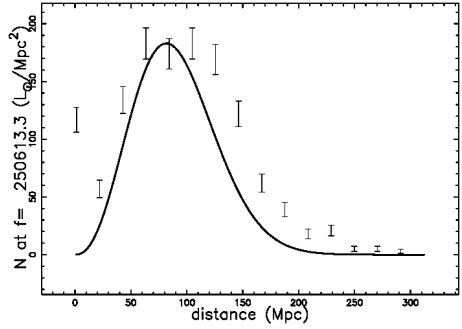}
\end{center}
\caption
{The galaxies of Cosmicflows-2 with
$ 12498.8
 \, L_{\sun}/Mpc^2  \leq
f \leq  488727.8 \, L_{\sun}/Mpc^2  $
are  organized by frequency versus
distance,  (empty circles);
the error bar is given by the square root of the frequency.
The maximum frequency of the observed galaxies is
at  $d=73.8 $\ Mpc.
The full line is the theoretical curve
generated by
$\frac{dN}{d\Omega dr df}$
as given by the application of the Schechter LF
which  is Equation~(\ref{nfunctionrschechter})
and the theoretical  maximum is at
$d=81.69$\ Mpc.
The parameters are
$L^*= 1.8 \,10^{10} L_{\sun}$   and
$\alpha$ =-0.83.
}
          \label{cosmicflows_maximum}%
    \end{figure}

\begin{figure*}
\begin{center}
\includegraphics[width=6cm]{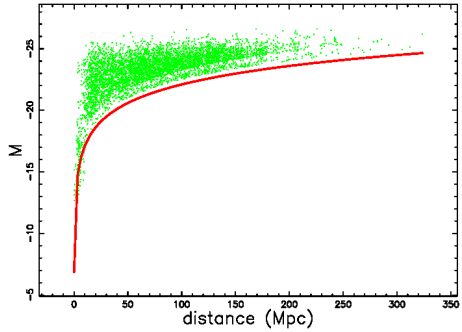}
\end{center}
\caption{The absolute magnitude $M$ of
4970 galaxies belonging to Cosmicflows-2
when the absolute bolometric 
magnitude  is  3.39 and
$H_{0}=74.4 \mathrm{\ km\ s}^{-1}\mathrm{\ Mpc}^{-1}$
(green points).
The upper theoretical curve, Equation~(\ref{distancemodulusgalaxies}),
is reported as a 
red thick line when $m$=15.39.
}
 \label{cosmicflows2_malmquist}%
 \end{figure*}

We now introduce the critical radius $r_{crit}$
\begin{equation}
r_{crit}=
\frac{1}{2}
\,{\frac{\sqrt{{\it L^* }}}{\sqrt{\pi}\sqrt{f}}}
\quad.
\end{equation}
Therefore
the joint distribution in distance
and  flux becomes
\begin{equation}
\frac{dN}{d\Omega dr df} =
\frac{1}{L^*} \,
4\,\pi\,{r}^{4}{\it \Phi^*}\, \left({\frac{{r}^{2}}{{{\it r_{crit}}}^{2}}}
 \right) ^{\alpha}{{\rm e}^{-{\frac{{r}^{2}}{{{\it r_{crit}}}^{2}}}}}
\quad.
\label{nfunctionrschechter_rcrit}
\end{equation}
The above  number of galaxies 
has a maximum at $r=r_{max}$:
\begin{equation}
r_{max}= \sqrt{2+\alpha}{\it r_{crit}}
\quad  ,
\label{rmaxrcrit}
\end{equation}
and the average  distance  of the galaxies,
${\langle r \rangle} $,
is
\begin{equation}
\langle r \rangle
={\frac{{\it r_{crit}}\,\Gamma \left( 3+\alpha \right) }
{\Gamma \left( \frac{5}{2} +\alpha \right) }}
\quad.
\label{raverageflux}
\end{equation}
Figure~\ref{cosmicflows_maximum}
presents the number of  galaxies   observed in Cosmicflows-2
as a function  of the distance  for  a given
window in  flux, as well as the theoretical curve.
The luminosity of a galaxy is produced and attenuated according 
to the electromagnetism.
We now deal with the mass of a galaxy.
A transformation of the luminosity of a galaxy, $L$,
by the mass, $\mathcal{M}$, is given by 
the 
 nonlinear  formula (9) in \cite{Cappellari_2006}
\begin{equation}
\frac{\mathcal{M}}{L} = 2.35
 \left(\frac{L}{10^{10} L_{\sun}}\right)^{0.32}
\label{eqnml}
\quad.
\end{equation}
We pay  careful attention  to the interval of existence in absolute 
magnitude versus distance for  
Cosmicflows-2 catalog,
see  Figure~\ref{cosmicflows2_malmquist}.

The progressive decrease of the interval of existence 
for the absolute magnitude is known
as Malmquist  bias \cite{Malmquist_1920}.
Conversely the mass  of  a galaxy  does not disappear  
with distance and therefore produce the gravitational field 
even if the galaxy is not visible due to 
the instrumental limitations.
\section{Gravitational forces}
\label{sec_gravitational}

\begin{figure}
\begin{center}
\includegraphics[width=6cm]{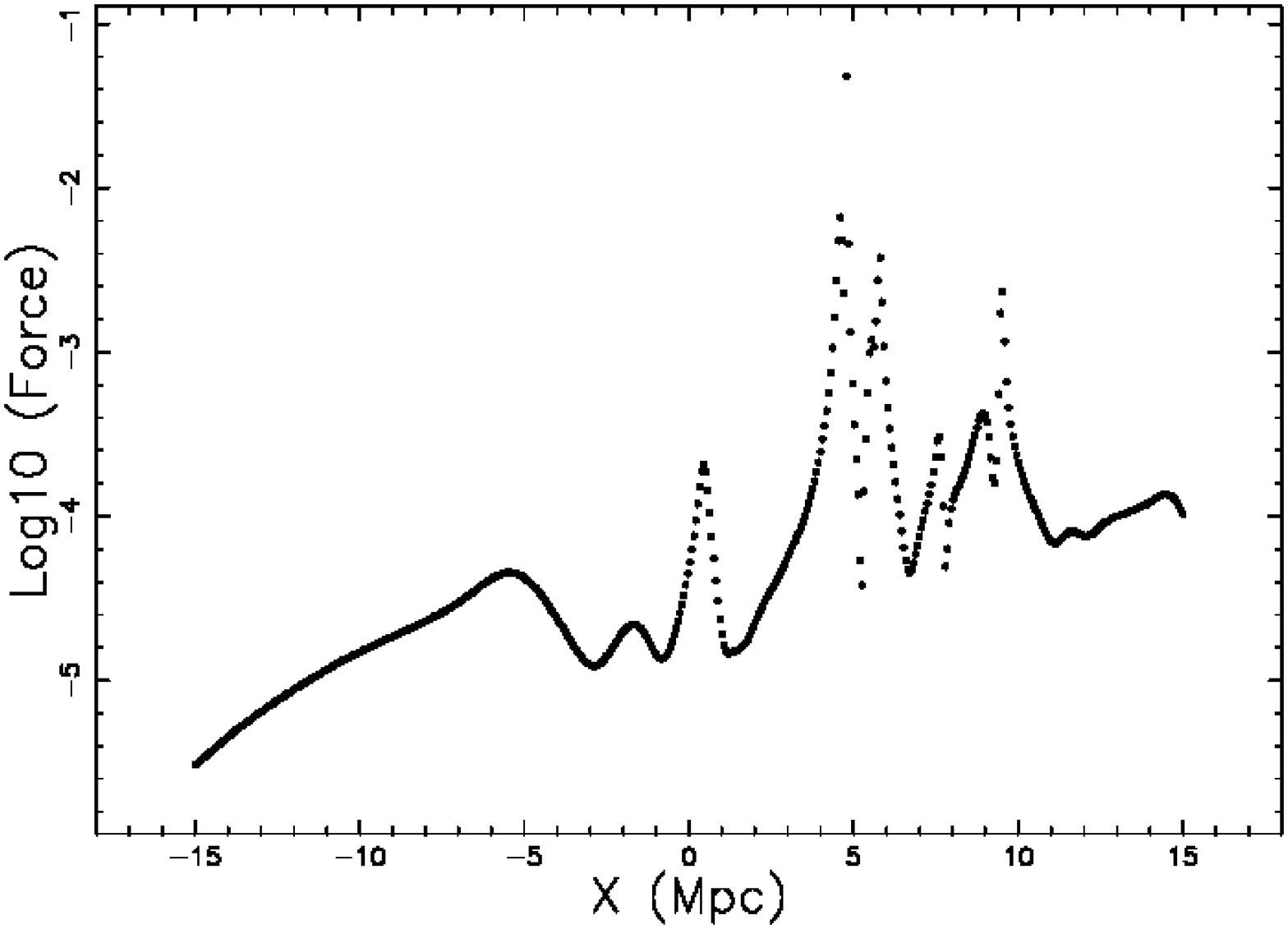}
\end{center}
\caption
{Cut-line  of the 
2D gravitational forces
expressed in $\frac{Mpc  M_{gal}}{yr8^2}$ 
(decimal logarithm).
}
          \label{cosmicflows2_2dfieldline}%
    \end{figure}

\begin{figure}
\begin{center}
\includegraphics[width=6cm]{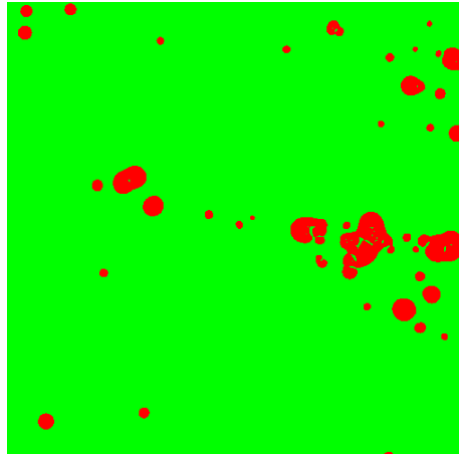}
\end{center}
\caption
{Color scheme for the values of the gravitational  field  
evaluated in  500$\times$500  points.
The red zone  
has  values of  gravitational field  in the  interval 
$\bigl [ 4.27 \frac{Mpc  M_{gal}}{yr8^2}  
,  0.00017
\frac{Mpc  M_{gal}}{yr8^2} \bigr ] $
(the zones near the galaxies)
and  the green zone 
has  values of  gravitational field in the  interval
$\bigl [ 0.00017 \frac{Mpc  M_{gal}}{yr8^2}  
,  1.7 \,10^{-8}
\frac{Mpc  M_{gal}}{yr8^2} \bigr ] $
( the zones of the voids). 
}
          \label{cosmicflows2_2dfield_2colors}%
    \end{figure}

The shell theorem, according to proposition LXX, theorem XXX,
in Principia \cite{Newton1687}
 states that 
"If to every point of a spherical surface there tend equal centripetal
forces decreasing
in the square of the distances from those points, I say, that a corpuscle
placed within
that superficies will not be attracted by those forces any way"
or  more simply
"The gravitational field inside a uniform spherical shell is zero" 
\cite{Arens_1990,Chandrasekhar1995,Borghi_2014}.
The galaxies are thought  to reside  on irregular
shells which are approximated by a given radius.
Due to the fact that the shapes of the above shells 
are neither exactly  spherical  nor uniformly populated by
galaxies we search for a low value of the gravitational field 
rather than the theoretical zero.
We therefore analyze a 2D box  with size of  30 $Mpc$ 
and masses as given by formula~\ref{eqnml}.
The forces  in every  point of the considered box   
are evaluated according 
to the Newtonian force where $G$, the Newtonian constant of 
gravitation, is 
\begin{equation}
G=4.49975  10^{-6}
\frac{Mpc^3}{M_{gal}yr8^2}
\quad ,
\end{equation}
where the length  is in $Mpc$,  
mass is in $M_{gal}$ which are 
$10^{11} M_{\sun}$  and  $yr8$
are  $10^8$ $yr$,
see more details in  \cite{Zaninetti2012i}.
The masses do  not disappear with distance 
and the above value of the box  
allows processing  a complete sample of galaxies.
Figure~\ref{cosmicflows2_2dfieldline} reports the values 
of the gravitational field as a cut  in  the middle 
of the  2D box.
Figure~\ref{cosmicflows2_2dfield_2colors} 
reports the values 
of the gravitational field
organized as a two color map.

\section{Conclusions}
\label{sec_conclusions}

The observed "great wall"  is theoretically 
explained by  a maximum 
in the number of galaxies as function of the distance
in Mpc, see Figure~\ref{cosmicflows_maximum}.
The two concepts of repeller and attractor, see \cite{Hoffman2017}
are not used in our analysis.
The presence  of voids approximated by spheres 
in the spatial distribution of galaxies 
allows to analyzing  the gravitational field
at the light of the shell theorem.
Due  to the facts that the number of galaxies 
is finite and the masses of galaxies are distributed 
according to a gamma probability density function
the gravitational field at the center 
of the voids turns out to be 
$\approx 2  \,10^{-8}
\frac{Mpc  M_{gal}}{yr8^2}$ rather than   zero.
\section*{Acknowledgments}

This research has made use of the VizieR catalogue access tool, CDS,
Strasbourg, France. 
\newpage  
\leftline{\bf References}
\providecommand{\newblock}{}

\end{document}